\documentclass[prd,preprint,showpacs,preprintnumbers,amsmath,amssymb]{revtex4}

\usepackage{graphicx}
\usepackage{dcolumn}
\usepackage{bm}

\begin{document}

\preprint{UTHET-07-0801}

\title{Quasi-normal modes of a black hole localized on a codimension-two brane of finite tension}

\author{Usama A. al-Binni}
\email{ualbinni@tennessee.edu}
\author{George Siopsis}%
 \email{siopsis@tennessee.edu}
\affiliation{%
Department of Physics and Astronomy,\\
The University of Tennessee, \\
Knoxville, TN 37996 - 1200, USA.\\
}%

\date{\today}

\begin{abstract}
We calculate the
quasi-normal modes of black holes localized on a 3-brane of finite tension
embedded in six-dimensional flat spacetime.
Using the WKB approximation, we obtain the frequencies of scalar and gravitational perturbations both analytically and numerically for arbitrary brane tension.
We thus extend earlier analytical results which were derived for a brane of small tension.
\end{abstract}

\maketitle

\section{Introduction}

Next to Hawking radiation, quasi-normal modes (QNMs) are an important signature to look for when searching for black holes.
They represent the damped oscillations under the effect of perturbation to the black-hole background. The oscillation occurs at characteristic frequencies that are complex numbers whose imaginary parts represent the damping.
QNMs have been studied extensively in the context of four-dimensional asymptotically flat spacetime (see \cite{KS,N} for general reviews of the earlier work on the subject).
However, when it comes to detection, neither signature has been observed yet and only indirect evidence has been used to infer the existence of astronomical black holes.
Nevertheless, gravitational wave observation to be hopefully achieved by projects like LIGO should offer a chance to detect QNMs in the near future.

Aside from astronomical interest, QNMs have been gaining attention recently for several reasons. The study of QNMs can shed light on the AdS/CFT correspondence (see, e.g., \cite{CM,HH}) and is relevant to the Barbero-Immirzi parameter \cite{bibbi1,bibbi2} of loop quantum gravity \cite{bibl1,bibl2,bibl3}.
In addition to these considerations, in brane-world models with large extra-dimensions \cite{ADD,ADD2,ADD3,bibRS} gravity is much stronger than observed, and the production and evaporation of mini black holes becomes a viable possibility at the LHC \cite{bibbh1,bibbh2,bibbh3,bibbh4,bibbh5} (see \cite{bibbhrev} for a recent review).
QNMs may then be observed at the LHC via the detection of the decay products of these mini black holes which will include Standard Model particles \cite{bibSM1,bibSM2,bibSM3,bibSM4,bibSM5}.

In most of the literature so far regarding black holes localized on branes, the tension of the brane has been assumed to be negligible.
This is hardly due to physical requirements, but rather due to lack of knowledge of solutions to the Einstein equations accommodating finite brane tension.
An interesting solution which does incorporate tension was constructed in \cite{KK}.
It describes a black hole residing in a tensional 3-brane embedded in six-dimensional locally flat spacetime.
A numerical study of its properties was performed in \cite{DKSS}. As the black hole evaporates,
it was found that the power emitted in the bulk decreases as the brane tension increases.
Thus for high tension the semi-classical approximation becomes more reliable.
Analytical results concerning the quasi-normal modes of scalar perturbations were derived in \cite{CWS}
to first order in the brane tension.

A major impediment to the analytical study of finite tension effects is the angular part of the wave equation as analytical solutions are not available for the metric of interest.
In our work we intend to close this gap and obtain analytical expressions which are valid for arbitrary brane tension.
In detail, in section \ref{sec2} we solve the angular wave equation analytically for finite tension, thus extending the first-order results of ref.~\cite{CWS}.
We also show that our analytical expressions are in agreement with the numerical results of ref.~\cite{DKSS}.
In section \ref{sec3} we calculate quasi-normal modes of scalar, electromagnetic and gravitational perturbations and discuss the dependence of their frequencies on the brane tension.
Finally in section \ref{sec4} we summarize our conclusions.


\section{\label{sec2}The Flow of Eigenvalues}

The metric of a black hole residing in a tensional 3-brane embedded in six-dimensional spacetime is given by \cite{KK}
\begin{equation}\label{eq1} \mathrm{d}s^{2}=-f(r)dt^{2}
+\frac{dr^{2}}{f(r)}+ r^2 d\Omega_4^2 \ \ , \ \ \ \
f(r)=1-\left(\frac{r_{h}}{r}\right)^{3} \end{equation}
where the angle element is
\begin{equation}\label{eq1a} d\Omega_4^2 = d\theta^{2}+\sin^{2}\theta
\left[ d\varphi^2+\sin^{2}\phi (d\chi ^{2}+
b^{2}\sin^{2}\chi d\psi ^{2}) \right] \ \ , \ \ \ \ 0 < b \le 1 \end{equation}
For $b=1$, this is the line element of the unit sphere $S^4$.
It corresponds to zero brane tension.
For non-vanishing tension, the parameter $b<1$ is a measure of the deficit angle about an axis parallel to the 3-brane in the angular direction $\psi$, such that the canonically normalized angle $\psi'=\psi/b$ runs over the interval $[0,2\pi/b]$.
It may be expressed as
\begin{equation}\label{eqb} b = 1 - \frac{\lambda}{4\pi M_*^4} \end{equation}
in terms of the brane tension $\lambda$ and the fundamental Planck constant of six-dimensional gravity $M_*$.

The radius of the horizon is given by
\begin{equation}\label{eq20}
r_h=\left(\frac{\mu}{b}\right)^{1/3} \ \ , \ \ \ \ \mu \equiv \frac{M_{BH}}{4\pi^2M_*^4}
\end{equation}
where $M_{BH}$ is the mass of the black hole.
Evidently, the tension of the brane modifies the radius of the horizon; it increases with increasing tension ($b\to 0$).


Let us now consider perturbations around the background
defined by the metric (\ref{eq1}).
First, we consider scalar perturbations which are simpler and then discuss electromagnetic and gravitational perturbations in the same manner.

The field equations of a scalar perturbation separate into a set of angular equations
\begin{equation}\label{eq10}
\begin{aligned}
&\frac{1}{\sin^3\theta}\frac{\mathrm{d}}{\mathrm{d}\theta}\left(\sin^3\theta\frac{\mathrm{d}}{\mathrm{d}\theta}\Theta(\theta)\right)+\left(\eta_3-\frac{\eta_2}{\sin^2\theta}\right)\Theta(\theta)
 = 0\\
&\frac{1}{\sin^2\varphi}\frac{\mathrm{d}}{\mathrm{d}\varphi}\left(\sin^2\varphi\frac{\mathrm{d}}{\mathrm{d}\varphi}\Phi(\varphi)\right)+\left(\eta_2-\frac{\eta_1}{\sin^2\varphi}\right)\Phi(\varphi)
 = 0\\
&\frac{1}{\sin\chi}\frac{\mathrm{d}}{\mathrm{d}\chi}\left(\sin\chi\frac{\mathrm{d}}{\mathrm{d}\chi}\Gamma(\chi)\right)+\left(\eta_1-\frac{\eta_0}{\sin^2\chi}\right)\Gamma(\chi)
 = 0\\
&\frac{1}{b^2}\frac{\mathrm{d^2}}{\mathrm{d}\psi^2}\Xi(\psi)+\eta_0\Xi(\psi)
 = 0\\
\end{aligned}
\end{equation}
and a radial equation that can be cast into the Schr\"odinger form \cite{bibIK}
\begin{equation}\label{eq11}
\frac{\mathrm{d}^2\Psi(r)}{\mathrm{d}r_*^2}+\left(\omega^2-V[r(r_*)]\right)\Psi(r)=0
\end{equation}
where $\Psi(r)=R(r)/r^2$, and $R(r)$ is the radial part of the
solution, in terms of the ``tortoise'' coordinate $r_*$ defined by
\begin{equation}\label{eq2}
\frac{\mathrm{d}r}{\mathrm{d}r_*} = f(r)
\end{equation}
which defines the ``tortoise'' coordinate $r_*$.
The effective potential
\begin{equation}\label{eqVsc} V(r) = f(r) \left( \frac{\eta_3+2}{r^2} + \frac{4r_h^3}{r^5} \right) \end{equation}
depends on the angular quantum number $\eta_3$ (eq.~(\ref{eq10})).
Therefore, in order to solve the eigenvalue problem (\ref{eq11}),
we ought to first solve the system of angular equations (\ref{eq10}).

The issue of finding the eigenvalues $\{\eta_i, i=0,\dots, 3 \}$ of the angular
equations (\ref{eq10}) was addressed numerically in \cite{DKSS} and
analytically (perturbatively to first order in the tension ($1-b$, see eq.~(\ref{eqb}))) in \cite{CWS}.

Here, we obtain exact analytical expressions of the eigenvalues as functions of the brane tension (arbitrary $b$).
To this end, we first consider the special case
\begin{equation}\label{eqbn} b = \frac{1}{N} \ \ , \ \ \ \ N\in\mathbb{N} \end{equation}
This set of special points includes high tension. Zero tension corresponds to the first point ($N=1$, $b=1$) and maximum tension ($b\to 0$) is an accumulation point of the set.
By restricting attention to the set (\ref{eqbn}), the eigenvalue problem reduces to one solvable by regular spherical harmonics.

The last angular equation (\ref{eq10}) is easily seen to yield the eigenvalues
\begin{equation}\label{eq3}
\eta_0=\frac{m}{b}=mN\ \ , \ \ \ \ m \in \mathbb{Z}
\end{equation}
For definiteness, let us restrict attention to $m\ge 0$. Negative values of $m$ can be treated similarly.

Working our way up, we note that for $b=1$ (regular sphere), the eigenvalue $\eta_1$ may be written as $\eta_1 = \ell_1 (\ell_1+1)$, where $\ell_1$ is an integer.
We also have $\eta_0 = m \le \ell_1$. It is easily seen that
for a general $b$ of the form (\ref{eqb}), $\eta_1$ is similarly given by
\begin{equation}\label{eq6}
\eta_1 =\lambda_1(\lambda_1+1) \ \ , \ \ \ \ \lambda_1 =\ell_1+(N-1)m=\ell_1+\left(\frac{1}{b}-1\right)m \end{equation}
Notice that $\lambda_1\le N\ell_1$ and so $\eta_0 = mN \le N\ell_1$, which is the maximum value of $\lambda_1$.

Going up to the next equation, we obtain the eigenvalue $\eta_2$ as
\begin{equation}\label{eq8}
\eta_2 =\lambda_2(\lambda_2+2) \ \ , \ \ \ \
\lambda_1 = \ell_2 + (N-1)m = \ell_2+\left(\frac{1}{b}-1\right)m
\end{equation}
reducing to the standard result on a three-sphere $\eta_2=\ell_2(\ell_2+2)$ with integer $\ell_2$ when $b=1$.

Finally, at the top of the ladder, we obtain the eigenvalue
\begin{equation}\label{eq9}
\eta_3 =\lambda_3(\lambda_3+3) \ \ , \ \ \ \
\lambda_3 = \ell_3 + (N-1)m =
\ell_3+\left(\frac{1}{b}-1\right)m
\end{equation}
reducing to the standard result on a four-sphere $\eta_3=\ell_3(\ell_3+2)$ with integer $\ell_3$ when $b=1$.

Although this result was derived for integral $1/b$ (eq.~(\ref{eqb})), it holds for
arbitrary $1/b$ as well ($b\in (0,1]$).
This may be seen by first generalizing to rational values of $b$, which may also be done in terms of ordinary spherical harmonics and then to arbitrary $b$ by noticing that the rational numbers form a dense subset of the interval $(0,1]$.
Alternatively, one may argue that the above expressions for the angular eigenvalues hold in general by analytic continuation.
We have verified their validity numerically using the Milne-spline method
\cite{YKMOTMWE} by least-square fitting the numerical results to the exact analytic expressions which resulted to a perfect fit.

\begin{figure}
  \includegraphics[width=6in]{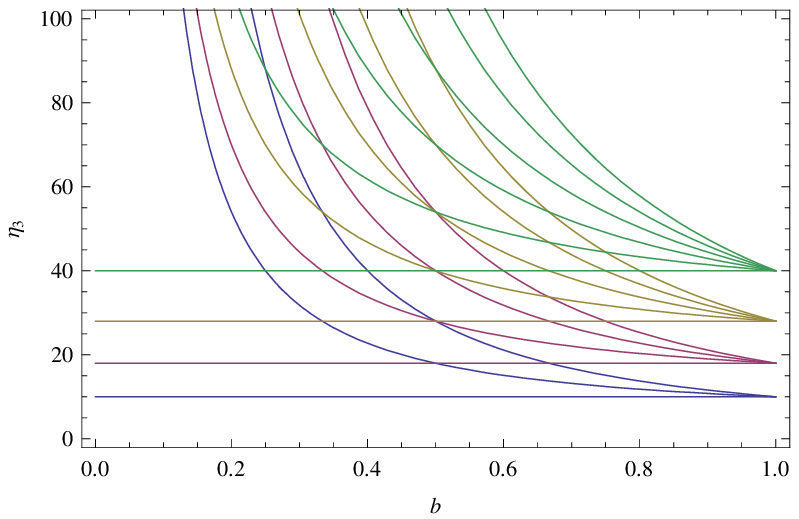}\\
  \caption{The flow of the angular eigenvalue $\eta_3$ (\ref{eq9}) with tension $b$, shown for different values of $\ell_3$ and $m$. Different colors belong to different values of $\ell_3=2,3,4,5$ (bottom to top).
}\label{fig1}
\end{figure}

Figure \ref{fig1} shows the flow of the eigenvalue $\eta_3$ with brane tension (as we vary $b$).
It also demonstrates that our analytical expression for $\eta_3$ (\ref{eq9}) reproduces the numerical
results of ref.~\cite{DKSS}.
Moreover, it is easily seen that (\ref{eq9}) agrees with the results of
\cite{CWS} to first order in the tension ($\frac{1}{b} - 1$).

Turning to other types of perturbation, we note that the angular equations differ slightly from the set of scalar equations (\ref{eq10}).
For vector and tensor perturbations, $\eta_3$ ought to be replaced by $\eta_3-1$ and $\eta_3 -2$, respectively.
The radial equations may still be written in the Schr\"odinger-like form (\ref{eq11}) but with different effective potentials \cite{bibIK,bibCCG,bibCHM}.
Scalar, electromagnetic, as well as vector and tensor gravitational perturbations
are described by the effective potential
\begin{equation}\label{eq16}
V_1(r)=f(r)\left(\frac{\eta_3}{r^2}+\frac{2}{r^2}+\frac{4(1-p^2)r_h^3}{r^5}\right)
\end{equation}
where $p$ takes on values depending on the type of perturbation: 0
for scalars and tensor gravitons, 2 for gravi-vectors, $1/2$ for
gauge vectors and $3/2$ for scalar reductions of bulk vectors.

On the other hand, the effective potential for scalar gravitational perturbations is given by \cite{bibIK,bibCCG}:
\begin{subequations}\label{eq17}
\begin{equation}
V_2(r)=f(r)\frac{Q(r)}{16r^2H(r)^2}
\end{equation}
where
\begin{equation}\label{eq18}
\begin{aligned}
Q(r)&=6400\left(\frac{r_h}{r}\right)^9+1920(\eta_3-4)\left(\frac{r_h}{r}\right)^6-1920(\eta_3-4)\left(\frac{r_h}{r}\right)^3\\
&+16(\eta_3-4)^3+96(\eta_3-4),\\
H(r)&=(\eta_3-4)+10\left(\frac{r_h}{r}\right)^3
\end{aligned}
\end{equation}
\end{subequations}
We shall use the form of the potential (\ref{eq16}) and (\ref{eq17}) to calculate the quasi-normal frequencies for the corresponding perturbations in the next section.

\section{\label{sec3}Quasi-normal frequencies}

Having obtained an exact analytical expression for $\eta_3$ (\ref{eq9}) valid for arbitrary brane tension ($0<b\le 1$), we may solve the radial equation (\ref{eq11}) with the effective potentials (\ref{eq16}) and (\ref{eq17})
for the various perturbations
to find the corresponding quasi-normal modes.
In \cite{CWS} the third-order WKB approximation was used to obtain the quasi-normal modes of scalar perturbations.
We will extend their results here by
including scalar, electromagnetic and gravitational perturbations and deriving expressions which are valid for arbitrary brane tension.

The third-order WKB formulas suitable for calculating black hole
quasi-normal modes were worked out in \cite{IW}.
The frequencies are given by
\begin{subequations}\label{12}
\begin{equation}\label{13}
\omega_n^2=\left[V_0+(-2V''_0)^{1/2}\Lambda\right]-i\alpha(-2V''_0)^{1/2}(1+\Omega)
\end{equation}
where
\begin{equation}\label{eq14}
  \Lambda  = \frac{1}
{{( - 2V''_0 )^{1/2} }}\left\{ \frac{1} {8}\left( {\frac{{V_0^{(4)}
}} {{V''_0 }}} \right)\left( {\frac{1}
{4} + \alpha } \right)
- \frac{1}
{{288}}\left( {\frac{{V'''_0 }}
{{V''_0 }}} \right)^2 (7 + 60\alpha ^2 ) \right\}
\end{equation}
\begin{equation}\label{eq15}
\begin{gathered}
  \Omega  = \frac{1}
{{( - 2V''_0 )^{1/2} }}\left\{ {\frac{5} {{6912}}\left(
{\frac{{V_0^{(4)} }}
{{V''_0 }}} \right)(77 + 188\alpha ^2 )
   - \left( {\frac{{{V'''}_0^2 V_0^{(4)} }}
{{{V''}_0^3 }}} \right)\frac{{51 + 100\alpha ^2 }} {{384}} } \right. \hfill \\
  \left. { + \left(
{\frac{{V_0^{(4)} }} {{V''_0 }}} \right)^2 \frac{{67 + 68\alpha ^2
}}
{{2304}}+ \left( {\frac{{V'''_0 V_0^{(5)} }}
{{{V''}_0^2 }}} \right)\frac{{19 + 28\alpha ^2 }} {{288}} - \left(
{\frac{{V_0^{(6)} }} {{V''_0 }}} \right)\frac{{5 + 4\alpha ^2 }}
{{288}}} \right\} \hfill \\
\end{gathered}
\end{equation}
and
\begin{equation}\label{eq19}
\alpha  = n + \frac{1} {2},\qquad V_0^{(s)}  = \left. {\frac{{d^s
V}} {{dr_*^s }}} \right|_{r_*  = r_* (r_p )}
\end{equation}
\end{subequations}
with $r_p$ being the peak of the potential.
Sample numerical results using the above formulas for the various types of perturbation (defined through the effective potentials $V_1$ (\ref{eq16}) and $V_2$ (\ref{eq17})) are shown in Tables \ref{tab:table2} and \ref{tab:table2a}.

\begin{table}
\caption{\label{tab:table2}3rd-order WKB approximation of the fundamental frequency.
First four columns correspond to the potential $V_1$ (\ref{eq16}); last column is for $V_2$ (\ref{eq17}). $\mu = 2$ and $\ell_3=5$.}
\scriptsize{
\begin{ruledtabular}
\begin{tabular}{llllll}
 $b$& $p=0$& $p=\frac{1}{2}$& $p=\frac{3}{2}$& $p=2$& $V_2$\\
 \multicolumn{6}{c}{$m=0$}\\
 \hline
1.0 & $2.94965-0.394141i$ & $2.9356-0.392963i$ & $2.82469-0.38449i$ & $2.73082-0.37905i$ & $2.60248-0.362449i$\\
0.9 & $2.84786-0.380539i$ & $2.83429-0.379402i$ & $2.72721-0.371221i$ & $2.63657-0.365969i$ & $2.51267-0.349941i$\\
0.7 & $2.61901-0.349959i$ & $2.60653-0.348913i$ & $2.50805-0.34139i$ & $2.4247-0.33656i$ & $2.31076-0.32182i$\\
0.5 & $2.34114-0.31283i$ & $2.32999-0.311895i$ & $2.24196-0.30517i$ & $2.16745-0.300852i$ & $2.06559-0.287676i$\\
0.3 & $1.97459-0.263851i$ & $1.96519-0.263062i$ & $1.89094-0.25739i$ & $1.8281-0.253749i$ & $1.74219-0.242635i$\\
0.1 & $1.36911-0.182944i$ & $1.36259-0.182397i$ & $1.3111-0.178464i$ & $1.26753-0.175939i$ & $1.20797-0.168234i$\\
0.01 & $0.635483-0.084915i$ & $0.632456-0.0846613i$ & $0.608561-0.0828358i$ & $0.588336-0.0816639i$ & $0.560688-0.0780873i$\\
0.001 & $0.294965-0.0394141i$ & $0.29356-0.0392963i$ & $0.282469-0.038449i$ & $0.273082-0.037905i$ & $0.260248-0.0362449i$\\
\hline\hline
 \multicolumn{6}{c}{$m=5$}\\
 \hline
1.0 & $2.94965-0.394141i$ & $2.9356-0.392963i$ & $2.82469-0.38449i$ & $2.73082-0.37905i$ & $2.60248-0.362449i$\\
0.9 & $3.09052-0.380255i$ & $3.07804-0.379296i$ & $2.9793-0.372292i$ & $2.89524-0.367484i$ & $2.7834-0.354961i$\\
0.7 & $3.47992-0.34922i$ & $3.47056-0.34864i$ & $3.39632-0.344283i$ & $3.33248-0.340972i$ & $3.25169-0.334445i$\\
0.5 & $4.1372-0.311811i$ & $4.13093-0.311522i$ & $4.08098-0.309295i$ & $4.0377-0.307481i$ & $3.98527-0.304652i$\\
0.3 & $5.51032-0.262764i$ & $5.50697-0.262668i$ & $5.48028-0.261908i$ & $5.45701-0.261261i$ & $5.42979-0.260422i$\\
0.1 & $10.8286-0.182101i$ & $10.8278-0.182093i$ & $10.8213-0.182027i$ & $10.8156-0.18197i$ & $10.809-0.181903i$\\
0.01 & $48.9429-0.0845181i$ & $48.9429-0.0845181i$ & $48.9426-0.0845178i$ & $48.9423-0.0845175i$ & $48.942-0.0845172i$\\
0.001 & $226.561-0.0392298i$ & $226.561-0.0392298i$ & $226.561-0.0392298i$ & $226.561-0.0392298i$ & $226.561-0.0392298i$\\
\end{tabular}
\end{ruledtabular}
}
\end{table}

\begin{table}
\caption{\label{tab:table2a}3rd-order WKB approximation of the first harmonic.
First four columns correspond to the potential $V_1$ (\ref{eq16}); last column is for $V_2$ (\ref{eq17}). $\mu = 2$ and $\ell_3=5$.}
\scriptsize{
\begin{ruledtabular}
\begin{tabular}{llllll}
 $b$& $p=0$& $p=\frac{1}{2}$& $p=\frac{3}{2}$& $p=2$& $V_2$\\
 \multicolumn{6}{c}{$m=0$}\\
 \hline
1.0 & $2.85074-1.1942i$ & $2.83667-1.19065i$ & $2.72427-1.16507i$ & $2.62697-1.14859i$ & $2.50456-1.09858i$\\
0.9 & $2.75236-1.15298i$ & $2.73878-1.14956i$ & $2.63025-1.12486i$ & $2.53632-1.10895i$ & $2.41813-1.06067i$\\
0.7 & $2.53118-1.06033i$ & $2.51869-1.05718i$ & $2.41889-1.03447i$ & $2.3325-1.01984i$ & $2.22381-0.975436i$\\
0.5 & $2.26263-0.947833i$ & $2.25147-0.945018i$ & $2.16225-0.924716i$ & $2.08503-0.911635i$ & $1.98787-0.871946i$\\
0.3 & $1.90838-0.799434i$ & $1.89896-0.797059i$ & $1.82372-0.779935i$ & $1.75858-0.768902i$ & $1.67663-0.735427i$\\
0.1 & $1.32319-0.554296i$ & $1.31667-0.55265i$ & $1.26449-0.540777i$ & $1.21933-0.533127i$ & $1.16251-0.509917i$\\
0.01 & $0.614173-0.257282i$ & $0.611143-0.256517i$ & $0.586926-0.251006i$ & $0.565964-0.247456i$ & $0.539591-0.236682i$\\
0.001 & $0.285074-0.11942i$ & $0.283667-0.119065i$ & $0.272427-0.116507i$ & $0.262697-0.114859i$ & $0.250456-0.109858i$\\
\hline\hline
 \multicolumn{6}{c}{$m=5$}\\
 \hline
1.0 & $2.85074-1.1942i$ & $2.83667-1.19065i$ & $2.72427-1.16507i$ & $2.62697-1.14859i$ & $2.50456-1.09858i$\\
0.9 & $3.00273-1.1503i$ & $2.99022-1.14741i$ & $2.8904-1.12629i$ & $2.804-1.11174i$ & $2.69594-1.07411i$\\
0.7 & $3.41427-1.05338i$ & $3.40488-1.05163i$ & $3.33014-1.03851i$ & $3.26539-1.02852i$ & $3.18572-1.00894i$\\
0.5 & $4.09324-0.938259i$ & $4.08695-0.937391i$ & $4.03683-0.930697i$ & $3.99329-0.92524i$ & $3.94102-0.91675i$\\
0.3 & $5.48691-0.789232i$ & $5.48356-0.788943i$ & $5.45683-0.786663i$ & $5.43352-0.78472i$ & $5.40629-0.7822i$\\
0.1 & $10.8229-0.546384i$ & $10.8221-0.546359i$ & $10.8156-0.546161i$ & $10.8098-0.545989i$ & $10.8033-0.545789i$\\
0.01 & $48.9427-0.253555i$ & $48.9426-0.253555i$ & $48.9423-0.253554i$ & $48.942-0.253553i$ & $48.9417-0.253552i$\\
0.001 & $226.561-0.117689i$ & $226.561-0.117689i$ & $226.561-0.117689i$ & $226.561-0.117689i$ & $226.561-0.117689i$\\
\end{tabular}
\end{ruledtabular}
}
\end{table}

We may also obtain explicit analytic expressions for the frequencies by expanding in inverse powers of the angular momentum quantum number $\ell_3$.
We obtain
\begin{equation}\label{eq22a}
\omega_n = \omega_n^{(0)} + \omega_n^{(1)}\ \frac{1}{\ell_3} + \mathcal{O} (1/\ell_3^2)
\end{equation}
where the leading contribution is given by
\begin{equation}\label{eq22b}
\omega_n^{(0)} = \frac{{\sqrt 3 }}
{{2^{2/3} 5^{5/6} } r_h}\ \left[ 2\lambda_3 +3 - i\sqrt{3} (2n+1) \right] \end{equation}
in terms of the radius of the horizon (\ref{eq20}) and the angular quantum number $\lambda_3$ (\ref{eq9}).
It is independent of the type of perturbation and reduces to the zero tension result (Schwarzschild black hole) \cite{K}
\begin{equation}\label{eq21}
  \omega_n  \approx \frac{1}
{2r_h}\left( {\frac{2}
{{D - 1}}} \right)^{\frac{1}
{{D - 3}}} \sqrt {\frac{{D - 3}}
{{D - 1}}}
\left[ 2\ell _3  +D- 3 - i\sqrt {D - 3} \left( {2n + 1} \right) \right]
\end{equation}
for $D=6$.

The first-order correction in (\ref{eq22a}) is given by
\begin{equation}\label{eq21a}
\omega_n^{(1)} = \frac{1}
{\sqrt{3}\ {2^{5/3} 5^{11/6} } r_h}\ \left[ 5-42(n+n^2)-48p^2 \right]
\end{equation}
for perturbations governed by the effective potential $V_1(r)$ (eq.~(\ref{eq16})), and by
\begin{equation}\label{eq21b}
\omega_n^{(1)} = \frac{{\sqrt 3 }}
{{2^{2/3} 5^{5/6} } r_h}\ \left[ -\frac{15}{2} +3(n+n^2) \right]
\end{equation}
for scalar gravitational perturbations governed by $V_2(r)$ (eq.~(\ref{eq17})).

It should be pointed out that the first-order WKB estimate \cite{SW}
\begin{equation}\label{28}
\omega ^2  = V_0  - i\alpha ( - 2V''_0 )^{1/2}
\end{equation}
suffices for the derivation of the above analytical asymptotic expressions.

\begin{figure}
  \includegraphics[width=6in]{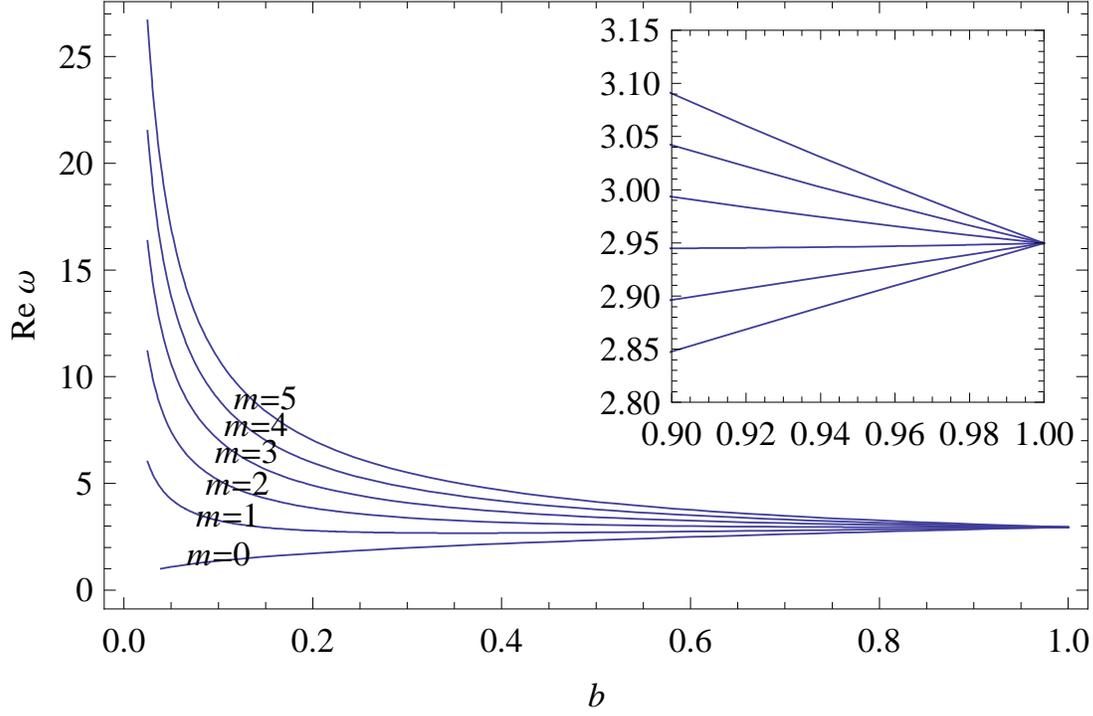}\\
  \caption{The real part of the fundamental frequency as a function of the tension $b$ for scalar perturbations with $\mu=2$, $\ell_3=5$ and $m=0,\dots,5$ (bottom to top). Detail showing degeneracy at zero tension ($b=1$).}\label{fig2}
\end{figure}

\begin{figure}
  \includegraphics[width=6in]{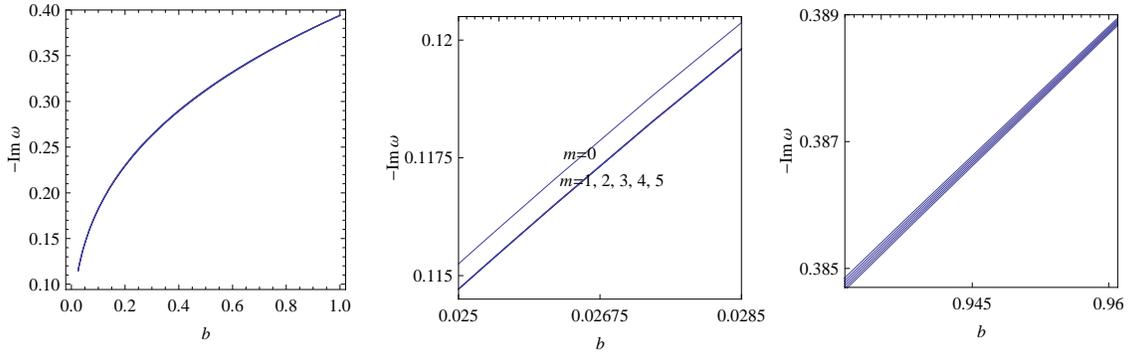}\\
  \caption{The imaginary part of the fundamental frequency for scalar perturbations with $\mu =2$ and $\ell_3=5$ and $M=2$. Detail showing splitting with $m$ at high ($b\to 0$) and low ($b\to 1$) tension.}\label{fig3}
\end{figure}

For the purpose of comparison with the results in
\cite{CWS}, we first look at scalar perturbations by setting $p=0$, $n=0$, $\ell_3=5$ and $\mu =2$ letting $m$ take the values 1 and 5 (for $m=0$ or $b=1$ the two approaches agree, because the angular eigenvalue $\eta_3$ (\ref{eq9}) agrees with its value on a regular sphere). This is shown in Table \ref{tab:table1} where we see that while the two cases agree perfectly on the imaginary part of the frequencies for any tension, the real part is over-estimated in the perturbative
calculation of \cite{CWS} at high tension (small $b$).
\begin{table}
\caption{\label{tab:table1}A comparison of
fundamental quasi-normal frequencies using {\em (a)} the exact expression (\ref{eq9}) for $\eta_3$ and {\em (b)} first-order approximate
eigenvalues calculated in \cite{CWS}. We set $\ell_3=5$, $\mu =2$.}
\begin{ruledtabular}
\begin{tabular}{ccc}
 $b$& {\em (a)} & {\em (b)} \\
 \multicolumn{3}{c}{$m=1$}\\
 \hline
 0.9&$2.89639-0.380476i$&$2.89863-0.380473i$ \\
 0.7&$2.79116- 0.349752 i$ &$2.8203- 0.349721 i$\\
 0.5&$2.70024- 0.312448i$&$2.82885- 0.312347$\\
 0.3&$2.68138- 0.263264i$&$3.15335- 0.263078i$\\
 0.1&$3.26011- 0.182229i$&$5.50665- 0.182137i$\\
 0.01&$10.2961- 0.0845195i$&$24.8883- 0.0845183i$\\
 \hline\hline
 \multicolumn{3}{c}{$m=5$}\\
 \hline
 0.9&$3.09052 - 0.380255i$&$3.09339 - 0.380252i$ \\
 0.7&$3.47992- 0.34922 i$&$3.51179- 0.349203 i$\\
 0.5&$4.1372- 0.311811i$ & $4.25276- 0.311786i$\\
 0.3&$5.51032- 0.262764i$&$5.84111- 0.262748i$\\
 0.1&$10.8286- 0.182101i$&$12.0048- 0.182099i$\\
 0.01&$48.9429- 0.0845181i$&$55.6375- 0.0845181i$\\
\end{tabular}
\end{ruledtabular}
\end{table}
Figures \ref{fig2} and \ref{fig3} show the
real and imaginary parts of the frequency as a function of $b$, respectively, for the chosen set of parameters.
The agreement on the imaginary parts can be traced to the asymptotic expression (\ref{eq22a})
which is valid for large $\ell_3$.
Unlike the real part, the imaginary part does not vary with tension except through the value of the radius of the horizon $\Im\omega \sim 1/r_h \sim b^{1/3}$ (eq.~(\ref{eq20})).
This also shows that the imaginary part of the frequency (damping) vanishes at high tension ($b\to 0$).

The real part, which represents the actual frequency of oscillation, also goes to zero at high tension ($\Re\omega \sim b^{1/3}$ as $b\to 0$) for $m=0$.
For all other values of the magnetic quantum number ($m\ne 0$), the real part diverges as $\Re\omega \sim b^{-2/3}$.
This level splitting is reminiscent of the Zeeman effect for an atom in a strong magnetic field
and is attributed to the spatial symmetry of the states.

At the other end of the spectrum (vanishing tension, $b\to 1$), frequencies become degenerate, as expected, because spherical symmetry is restored in the absence of a brane
(Schwarzschild solution).

\begin{figure}
  \includegraphics[width=6in]{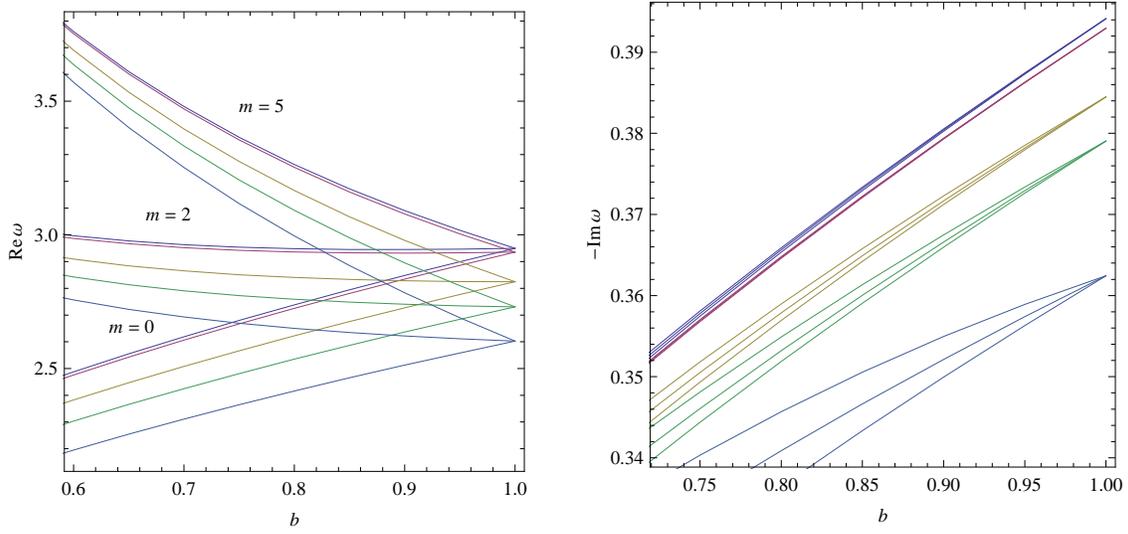}\\
  \caption{Real and imaginary parts of the fundamental frequency as a function of the tension $b$ for different perturbations with $\mu =2$, $\ell_3=5$, $m=0,2,5$. Top to bottom: $p=0,\frac{1}{2},\frac{3}{2},2$ (\ref{eq16}) and scalar gravitational (\ref{eq17}).} \label{fig4}
\end{figure}

\begin{figure}
  \includegraphics[width=6in]{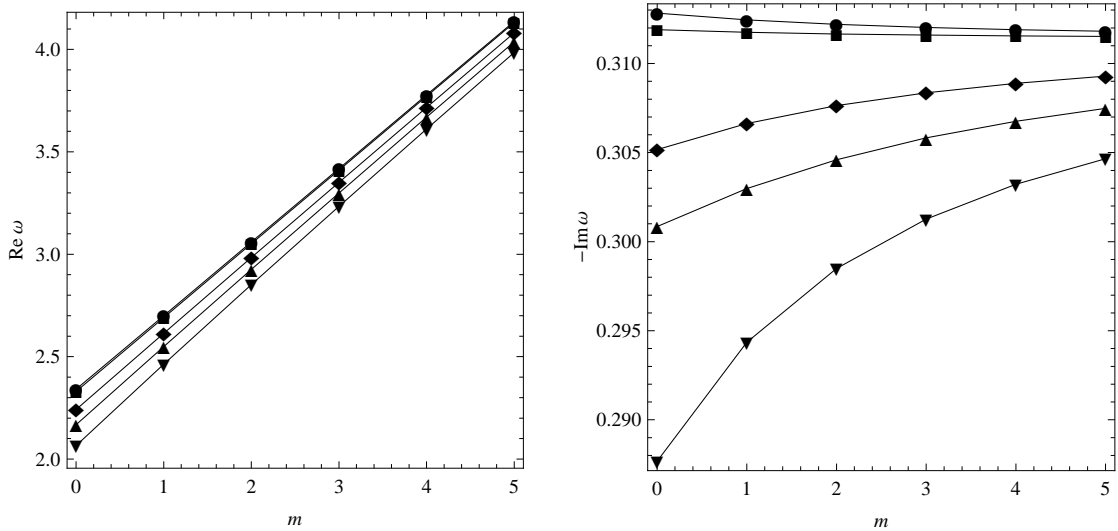}\\
  \caption{Dependence of the fundamental frequency on $m$ for $b=0.5$ and $\ell_3=5$. $p$ (\ref{eq16}) increases from top to bottom; scalar gravitational (\ref{eq17}) is represented by inverted triangles.}\label{fig5}
\end{figure}

Figure \ref{fig4} shows the dependence of the fundamental frequency on
the type of perturbation.
The graphs are in agreement with the analytic result (\ref{eq22a}) which shows that the dependence on
the type of perturbation is a next-to-leading-order effect (expanding in inverse powers of the angular momentum quantum number).
Similar results hold for higher harmonics as well.
In figure \ref{fig5}, in addition to the type of perturbation, the magnetic quantum number $m$ is also allowed to vary.
For the potential $V_1$ (\ref{eq16}), the real part decreases as the parameter $p$ labeling the type of perturbation decreases and the real part of the frequency for $V_2$ (\ref{eq17}) is lower than the rest.
The imaginary part of the frequency exhibits a similar behavior for the potential $V_1$ with the imaginary part of the frequency of $V_2$ being second-to-lowest.

The numerical accuracy of the analytic expression for quasi-normal frequencies (eqs.~(\ref{eq22a}), (\ref{eq22b}) and (\ref{eq21a})) is explored in Tables \ref{tab:table3} and \ref{tab:table3a}.
Evidently, the agreement is fairly good throughout the range of $b$, for both high and low brane tension.
The agreement becomes better as the angular momentum quantum number $\ell_3$ increases,
as expected.
The tables only show results for the fundamental frequency, however similar agreement is obtained for higher higher harmonics.

\begin{table}
\caption{\label{tab:table3}Comparison
between the $\mathcal{O} (1/\ell_3^2)$ analytic expression (\ref{eq22a}) and the WKB approximation of fundamental quasi-normal frequencies for the potential $V_1$ (\ref{eq16}).
}
\begin{ruledtabular}
\begin{tabular}{llll}
 $b$& analytic & 1st-order WKB& 3rd-order WKB\\
 \multicolumn{4}{c}{$\ell_3=5$, $m=0$, $p=\frac{1}{2}$}\\
\hline
1.0 & $2.93385-0.392298i$ & $3.00359-0.392395i$ & $2.9356-0.392963i$\\
0.9 & $2.8326-0.37876i$ & $2.89994-0.378853i$ & $2.83429-0.379402i$\\
0.7 & $2.60497-0.348323i$ & $2.6669-0.348409i$ & $2.60653-0.348913i$\\
0.5 & $2.32859-0.311367i$ & $2.38395-0.311444i$ & $2.32999-0.311895i$\\
0.3 & $1.96401-0.262617i$ & $2.01071-0.262682i$ & $1.96519-0.263062i$\\
0.1 & $1.36177-0.182089i$ & $1.39415-0.182133i$ & $1.36259-0.182397i$\\
0.01 & $0.632078-0.0845181i$ & $0.647105-0.0845389i$ & $0.632456-0.0846613i$\\
0.001 & $0.293385-0.0392298i$ & $0.300359-0.0392395i$ & $0.29356-0.0392963i$\\
\hline\hline
\multicolumn{4}{c}{$\ell_3=10$, $m=5$, $p=\frac{3}{2}$}\\
 \hline
1.0 & $5.13159-0.392298i$ & $5.17978-0.389449i$ & $5.14172-0.389687i$\\
0.9 & $5.19747-0.37876i$ & $5.24531-0.376253i$ & $5.21026-0.376461i$\\
0.7 & $5.41823-0.348323i$ & $5.46515-0.346517i$ & $5.43666-0.346666i$\\
0.5 & $5.87062-0.311367i$ & $5.91599-0.31026i$ & $5.89493-0.31035i$\\
0.3 & $6.9731-0.262617i$ & $7.01534-0.262142i$ & $7.00269-0.26218i$\\
0.1 & $11.8435-0.182089i$ & $11.8768-0.182033i$ & $11.8732-0.182038i$\\
0.01 & $49.4141-0.0845181i$ & $49.4307-0.0845178i$ & $49.4306-0.0845178i$\\
0.001 & $226.78-0.0392298i$ & $226.788-0.0392298i$ & $226.788-0.0392298i$\\
\end{tabular}
\end{ruledtabular}
\end{table}
\begin{table}
\caption{\label{tab:table3a}Comparison
between the $\mathcal{O} (1/\ell_3^2)$ analytic expression (\ref{eq22a}) and the WKB approximation of fundamental quasi-normal frequencies for the potential $V_2$ (\ref{eq17}).
}
\begin{ruledtabular}
\begin{tabular}{llll}
 $b$& analytic & 1st-order WKB& 3rd-order WKB\\
\multicolumn{4}{c}{$\ell_3=10$, $m=0$}\\
 \hline
1.0 & $5.03948-0.392298i$ & $5.05884-0.383546i$ & $5.02112-0.383837i$\\
0.9 & $4.86556-0.37876i$ & $4.88426-0.370309i$ & $4.84784-0.370591i$\\
0.7 & $4.47457-0.348323i$ & $4.49176-0.340552i$ & $4.45828-0.34081i$\\
0.5 & $3.99984-0.311367i$ & $4.0152-0.304421i$ & $3.98527-0.304652i$\\
0.3 & $3.37359-0.262617i$ & $3.38655-0.256758i$ & $3.36131-0.256953i$\\
0.1 & $2.33912-0.182089i$ & $2.34811-0.178026i$ & $2.3306-0.178161i$\\
0.01 & $1.08572-0.0845181i$ & $1.08989-0.0826325i$ & $1.08177-0.0826952i$\\
0.001 & $0.503948-0.0392298i$ & $0.505884-0.0383546i$ & $0.502112-0.0383837i$\\
\hline\hline
\multicolumn{4}{c}{$\ell_3=10$, $m=5$}\\
 \hline
1.0 & $5.03948-0.392298i$ & $5.05884-0.383546i$ & $5.02112-0.383837i$\\
0.9 & $5.10854-0.37876i$ & $5.13417-0.371112i$ & $5.09939-0.371363i$\\
0.7 & $5.33645-0.348323i$ & $5.37525-0.342898i$ & $5.34693-0.34307i$\\
0.5 & $5.79752-0.311367i$ & $5.84989-0.308097i$ & $5.82891-0.308197i$\\
0.3 & $6.91144-0.262617i$ & $6.97585-0.261239i$ & $6.96322-0.261279i$\\
0.1 & $11.8007-0.182089i$ & $11.8656-0.18193i$ & $11.862-0.181934i$\\
0.01 & $49.3943-0.0845181i$ & $49.4302-0.0845172i$ & $49.43-0.0845172i$\\
0.001 & $226.771-0.0392298i$ & $226.788-0.0392298i$ & $226.788-0.0392298i$\\
\end{tabular}
\end{ruledtabular}
\end{table}

\section{\label{sec4} Conclusions}

We analyzed the effect of brane tension on the low-lying quasi-normal modes
of a black hole localized on a tense 3-brane in six flat dimensions \cite{KK}.
In the presence of the brane, the symmetry is no longer spherical resulting
in angular wave equations for which an analytic solution is not available.
The properties of the black hole as the brane tension varies were studied
numerically \cite{DKSS} as well as analytically to first order in the tension \cite{CWS}.
We presented an analytic solution to the angular eigenvalue problem for arbitrary tension (parametrized by $b\in (0,1]$ which is related to the deficit angle which appears in the presence of the brane).
This enabled us to obtain explicit analytic expressions for quasi-normal modes of various types of perturbation.
We compared the analytic expressions to numerical results using 3rd-order WKB approximation and found good agreement.
We discussed the effect of brane tension on the frequencies. For low tension ($b\to 1$), we recovered the results of spherical geometry (Schwarzschild black hole).
For high tension ($b\to 0$), the imaginary part of all frequencies (damping) was seen to vanish as $b^{1/3}$ (being inversely proportional to the radius of the horizon).
The real part (representing the actual oscillation) also vanished for $m=0$ but
diverged as $b^{-2/3}$ for $m\ne 0$ similarly to the Zeeman effect for an atom
in a strong magnetic field.

Understanding the effects of brane tension is important in relation to the
detection of black holes at the LHC.
This is a difficult study due to the lack of solutions to the Einstein equations
describing black holes localized in branes of finite tension.
As far as we know, ours is the first study of analytic properties of perturbations of such black holes.
It would be interesting to further analyze the properties of these black holes including gray-body factors, the distribution of radiation, etc.,
and calculate their effect in the LHC environment.
Work in this direction is in progress.

\section*{Acknowledgment}

Research supported in part by the Department of
Energy under grant DE-FG05-91ER40627.

\bibliography{paper}
\newpage

\end{document}